\documentclass[aps,prl,twocolumn,showpacs,floatfix,superscriptaddress]{revtex4}

\usepackage{amsmath,amssymb}
\usepackage[dvips]{graphics}
\usepackage{epsfig}

\newcommand{\varA}{{\mathcal{A}}}
\newcommand{\varE}{{\mathcal{E}}}
\newcommand{\varH}{{\mathcal{H}}}

\newcommand{\bfk}{{\mathbf{k}}}
\newcommand{\bfS}{{\mathbf{S}}}
\newcommand{\bfsigma}{{\boldsymbol{\sigma}}}

\newcommand{\E}{{\rm e}}

\newcommand{\up}{{\uparrow}}
\newcommand{\down}{{\downarrow}}

\newcommand{\ode}[3][]{\frac{d^{#1}{#2}}{d{#3}^{#1}}}

\newcommand{\eqnref}[1]{Eq.~(\ref{#1})}

\newcommand{\figref}[1]{Fig.~\ref{#1}}
\newcommand{\Figref}[1]{Figure~\ref{#1}}

\begin{document}
\title{Josephson Effect through an Isotropic Magnetic Molecule}
\author{Minchul Lee}
\affiliation{Centre de Physique Th\'eorique, UMR6207, Case 907, Luminy, 13288 Marseille Cedex 9, France}
\author{Thibaut Jonckheere}
\affiliation{Centre de Physique Th\'eorique, UMR6207, Case 907, Luminy, 13288 Marseille Cedex 9, France}
\author{Thierry Martin}
\affiliation{Centre de Physique Th\'eorique, UMR6207, Case 907, Luminy, 13288 Marseille Cedex 9, France}
\affiliation{Universit\'e de la M\'editerran\'ee, 13288 Marseille Cedex 9, France}

\begin{abstract}
  We investigate the Josephson effect through a molecular quantum dot magnet
  connected to superconducting leads. The molecule contains a magnetic atom,
  whose spin is assumed to be isotropic.  It is coupled to the electron spin on
  the dot via exchange coupling.  Using the numerical renormalization group
  method we calculate the Andreev levels and the supercurrent and examine
  intertwined effect of the exchange coupling, Kondo correlation, and
  superconductivity on the current. Exchange coupling typically suppresses the
  Kondo correlation so that the system undergoes a phase transition from 0 to
  $\pi$ state as the modulus of exchange coupling increases. Antiferromagnetic
  coupling is found to drive exotic transitions: the reentrance to the $\pi$
  state for a small superconducting gap and the restoration of 0 state for
  large antiferromagnetic exchange coupling. We suggest that the asymmetric
  dependence of supercurrent on the exchange coupling could be used as to
  detect its sign in experiments.
\end{abstract}

\maketitle

Molecular spintronics \cite{Rocha05} aims at exploring spin-dependent
electronic transport through molecules with intrinsic degrees of freedom such
as spin, connected to leads of various
nature. 
On the theoretical and experimental side, recent advances have concerned both
coherent \cite{Romeike1} and incoherent \cite{Elste,Romeike2,Heersche06}
transport through these molecular quantum dot magnets (MQDM).  They consist of
a magnetic molecule with either a large \cite{Sessoli93} or a small anisotropy,
as is the case for a endofullerene molecule \cite{Grose}.

Here, we provide a nonperturbative computation of the low temperature transport
properties of a MQDM connected to superconducting leads using a numeral
renormalization group (NRG) approach.  The Josephson current allows a diagnosis
of the interaction between the intrinsic spin of the molecule, its itinerant
electron spin, and the polarization of the leads.  It has been known for some
time \cite{Shiba69,Glazman89,Spivak91,Rozhkov}, and recently analyzed in
experiments \cite{vanDam06}, that a quantum dot sandwiched between
superconducting leads can show a $\pi$ junction behavior \cite{ExpPiJunction}.
At the same time, a quantum dot connected to leads at low enough temperatures
exhibits the Kondo effect \cite{golhaber_gordon}.  It was shown
\cite{Glazman89,Choi04,Siano04} that with superconducting leads, at low
temperature the 0 junction state of the Josephson current is restored when the
Kondo temperature exceeds the superconducting gap.  The stability of this Kondo
phase is put in question in the presence of additional spin degrees for freedom
\cite{Bergeret06} which may compete with Kondo screening.  Here the Josephson
current flows through an isotropic MQDM which can describe a endofullerene
molecule \cite{Kasumov05}.  The electron spin in the quantum dot and the
magnetic ion inside it interact via an exchange coupling \cite{Elste}.  We
calculate the Andreev level (AL) spectrum and the supercurrent and determine
the spin of the ground state. We find that the exchange coupling typically
suppresses the Kondo effect and drives a transition from 0 to $\pi$
state. Moreover, antiferromagnetic coupling is found to drive exotic
transitions: the reentrance to $\pi$ state for small superconducting gap and
the restoration of 0 state for large $J$.

\begin{figure}[!t]
  \centering
  \includegraphics[width=6.5cm]{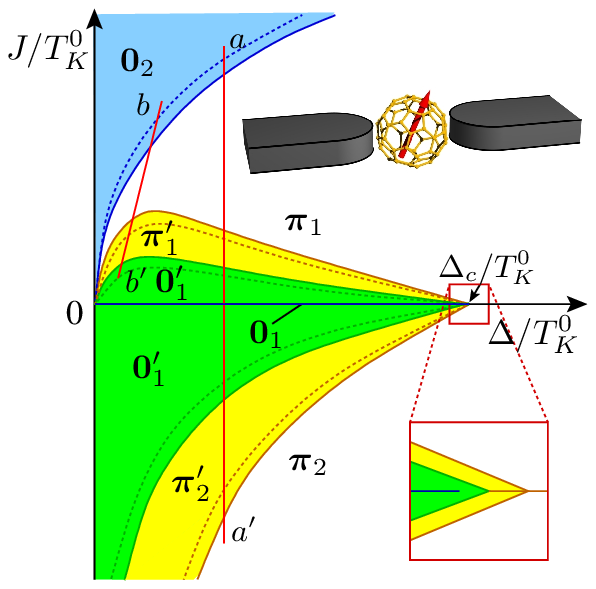}
  \caption{(color online) Schematic phase diagram of a MQDM superconducting
    junction system [see the upper inset] indicating the 0, 0' (blue), $\pi'$
    (green), and $\pi$ regions. Each region is divided into two subregions
    according to the ground-state spin: $S$ and $S{-}1/2$ for $0_1^{(\prime)}$
    and $0_2$ regions and $S{-}1/2$ and $S{+}1/2$ for $\pi_1^{(\prime)}$ and
    $\pi_2^{(\prime)}$ regions, respectively. Note that the $0_1$ state exists
    only along the line $J{=}0$ [see the lower inset]. For larger molecular
    spin $S' {>} S$ (see the dotted lines), the phase boundaries between $0_1$
    and $\pi_{1/2}$ are shifted toward smaller $|J|$, and one between $0_2$ and
    $\pi_1$ moves toward larger $J$.}
  \label{fig:1}
\end{figure}

The MQDM connected to two $s$-wave superconducting leads (inset of
\figref{fig:1}) is modeled by a single-impurity Anderson model: $\varH =
\varH_{\rm M} + \varH_{\rm L} + \varH_{\rm T}$, where
\begin{align}
  \varH_{\rm M}
  & =
  \epsilon_0 n + U n_\up n_\down + J \bfS\cdot\bfS_\E
  \\
  \varH_{\rm L}
  & =
  \sum_{\ell\bfk}
  \left[
    \epsilon_{\bfk} n_{\ell\bfk}
    -
    \left(
      \Delta\, e^{i\phi_\ell} c_{\ell\bfk\up}^\dag c_{\ell-\bfk\down}^\dag
      + (h.c.)
    \right)
  \right]
  \\
  \varH_{\rm T}
  & =
  \sum_{\ell\bfk\mu}
  \left[t\, d_\mu^\dag c_{\ell\bfk\mu} + (h.c.)\right].
\end{align}
Here $c_{\ell\bfk\mu}$ ($d_\mu$) destroys an electron with energy
$\epsilon_{\bfk}$,
and spin $\mu$ on lead $\ell=L,R$ (on the carbon cell); $n_{\ell\bfk}$
and $n$
are occupation operators for the leads and the cell.  The single-particle
energy $\epsilon_0$ can be tuned by gate voltages.  $J$ denotes the exchange
energy between the ion spin $\bfS$ and the electron spin $\bfS_\E = \frac12
\sum_{\mu\mu'} d_\mu^\dag \bfsigma_{\mu\mu'} d_{\mu'}$.
$\Delta$ is the superconducting gap.  Except for the finite phase difference
$\phi=\phi_L-\phi_R$, the leads are identical and their coupling to the MQDM is
symmetric.  The hybridization between the molecule and the leads is well
characterized by a tunneling rate $\Gamma = \pi \rho_0 |t|^2$, where $\rho_0$
is the density of states of the leads at the Fermi energy. As we are interested
in the low temperature behavior, we concentrate for the most part on the Kondo
regime with a localized level $-\epsilon_0 \gg \Gamma$ with large charging
energy $U\gg|\epsilon_0|$. Specifically, we choose $\epsilon_0 = -0.1 D$ (the
band width $D$ is taken as a unit of energy), $\Gamma = 0.01 D$, and $U=\infty$
and introduce the bare Kondo temperature $T_K^0 = \sqrt{D\Gamma/2} \exp
\left[\frac{\pi\epsilon_0}{2\Gamma}\left(1 + \frac{\epsilon_0}{U}\right)
\right]$ (at $J=\Delta=0$). The energy spectrum is found with the NRG method
\cite{NRG} extended to superconducting leads \cite{Yoshioka00,Choi04}. Within
the NRG method, the supercurrent is directly obtained by evaluating the
expectation value of the current operator \cite{Choi04}.

\figref{fig:1} shows the phase diagram of our system, which constitutes the
main result. The junction property switches between 0 and $\pi$ state,
depending on the strengths of $J$ and $\Delta$ with respect to $T_K^0$. For
$J=0$, the system undergoes the Kondo-driven phase transition
\cite{Glazman89,Choi04,Siano04}: The ground-state wave function is of spin
singlet kind for $\Delta < \Delta_c \approx 1.84\, T_K^0$ and of spin doublet
for $\Delta > \Delta_c$. In the strong coupling limit ($\Delta < \Delta_c$)
Kondo correlations screen out the localized spin and Cooper pairs tunnel
through the Kondo resonance state, resulting in a 0-junction \cite{
  Choi04,Siano04}. In the weak coupling limit ($\Delta > \Delta_c$), strong
superconductivity in the leads leaves the local spin unscreened and the
tunneling of Cooper pairs subject to strong Coulomb interaction acquires an
additional phase $\pi$, making a $\pi$-junction
\cite{Shiba69,Glazman89,Spivak91,Choi04,Siano04}. It is also found
\cite{Choi04} that the transition is $\phi$-dependent so that a narrow region
of the intermediate states $0'$ and $\pi'$ exists; see the enlarged view in
\figref{fig:1}.

\begin{figure}[!t]
  \centering
  \includegraphics[width=8cm]{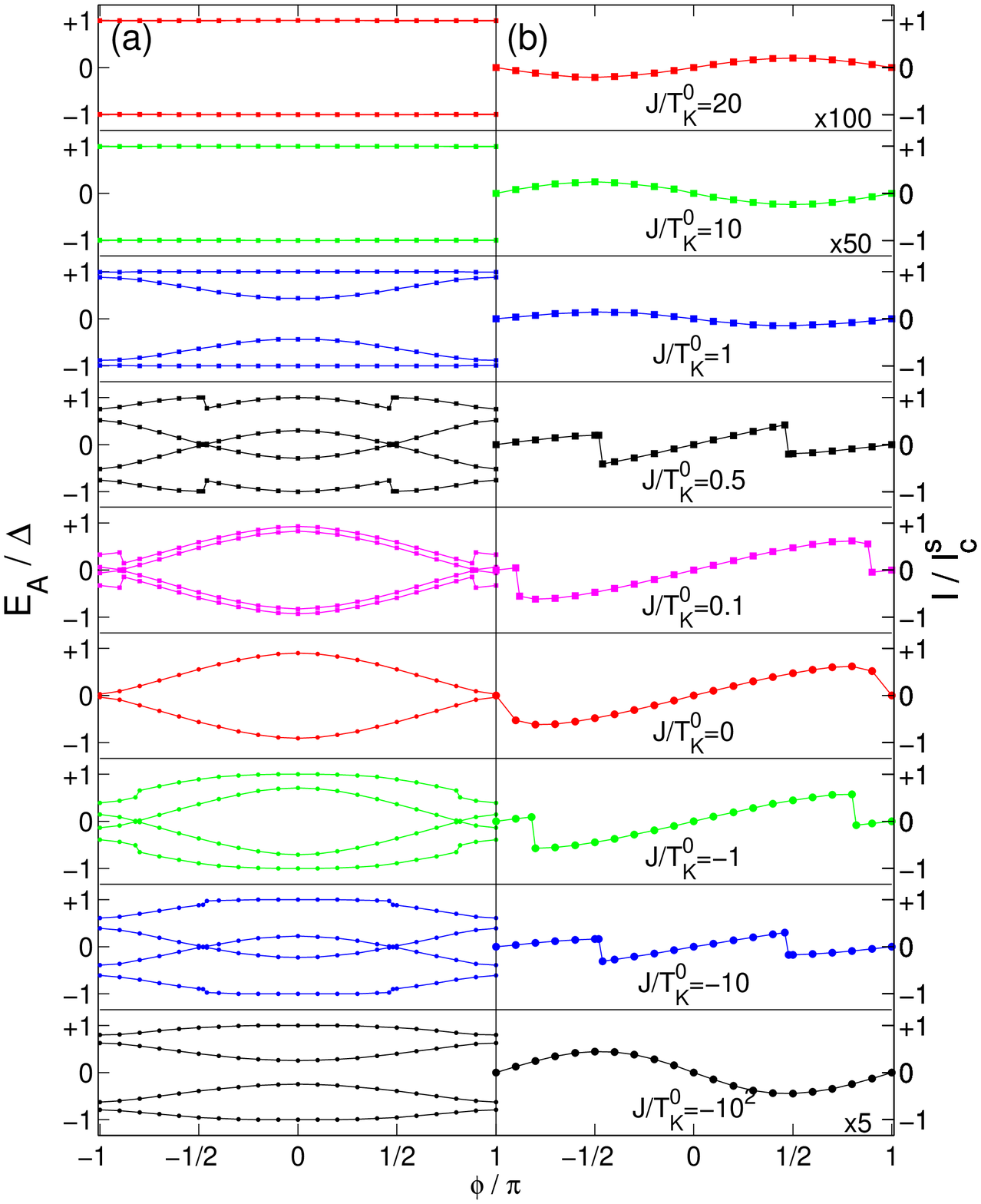}
  \caption{(color online) (a) ALs in units of $\Delta$ and (b) supercurrents
    $I$ in units of $I_c^s \equiv e\Delta/\hbar$ as functions of $\phi$ in the
    strong coupling limit ($\Delta/T_K^0 = 0.1$) for various values of
    $J/T_K^0$: see the line $aa'$ in \figref{fig:1}. Here the ion spin $S$ is
    set to $1/2$.}
  \label{fig:2}
\end{figure}

\begin{figure}[!t]
  \centering
  \includegraphics[width=4.25cm]{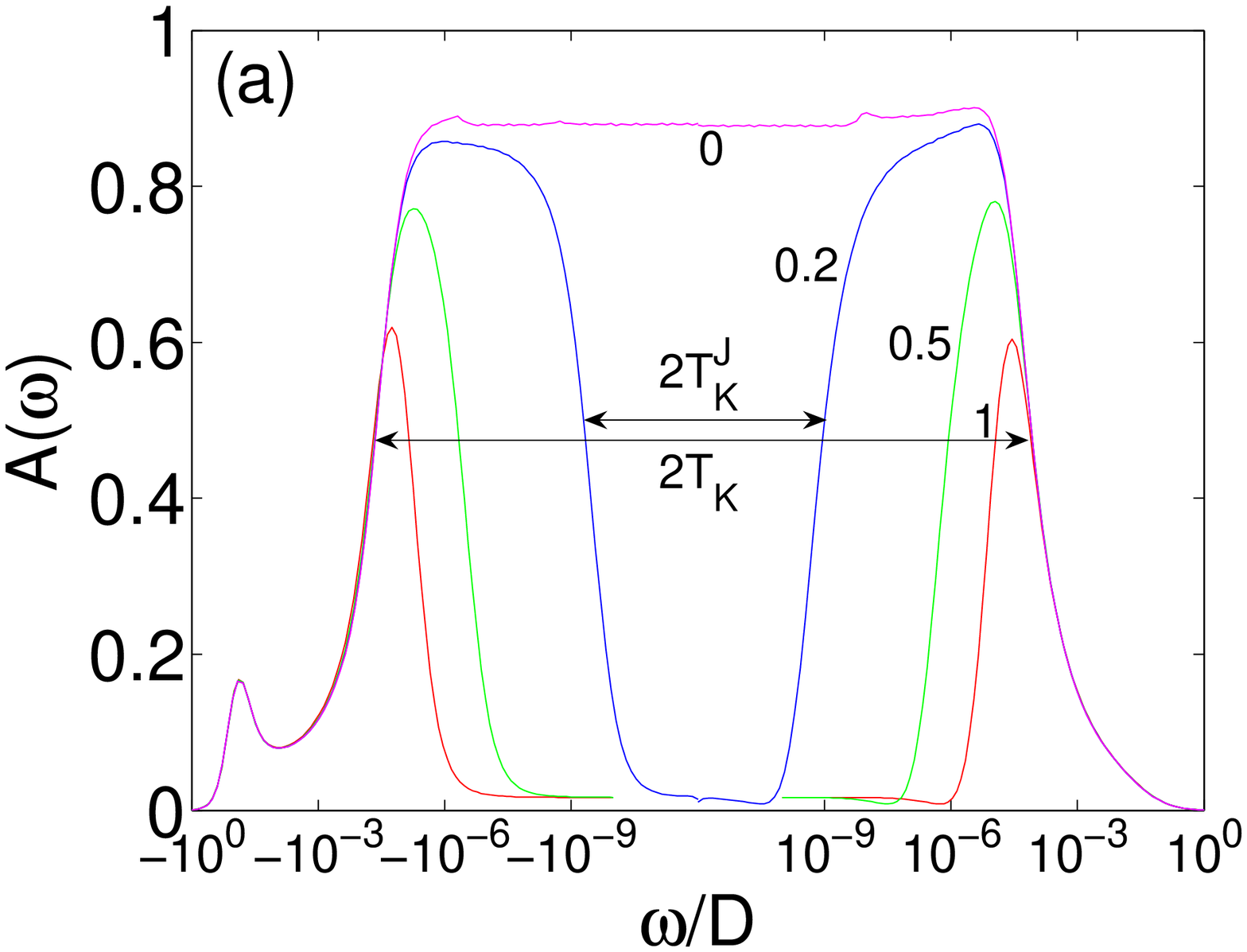}%
  \includegraphics[width=4.25cm]{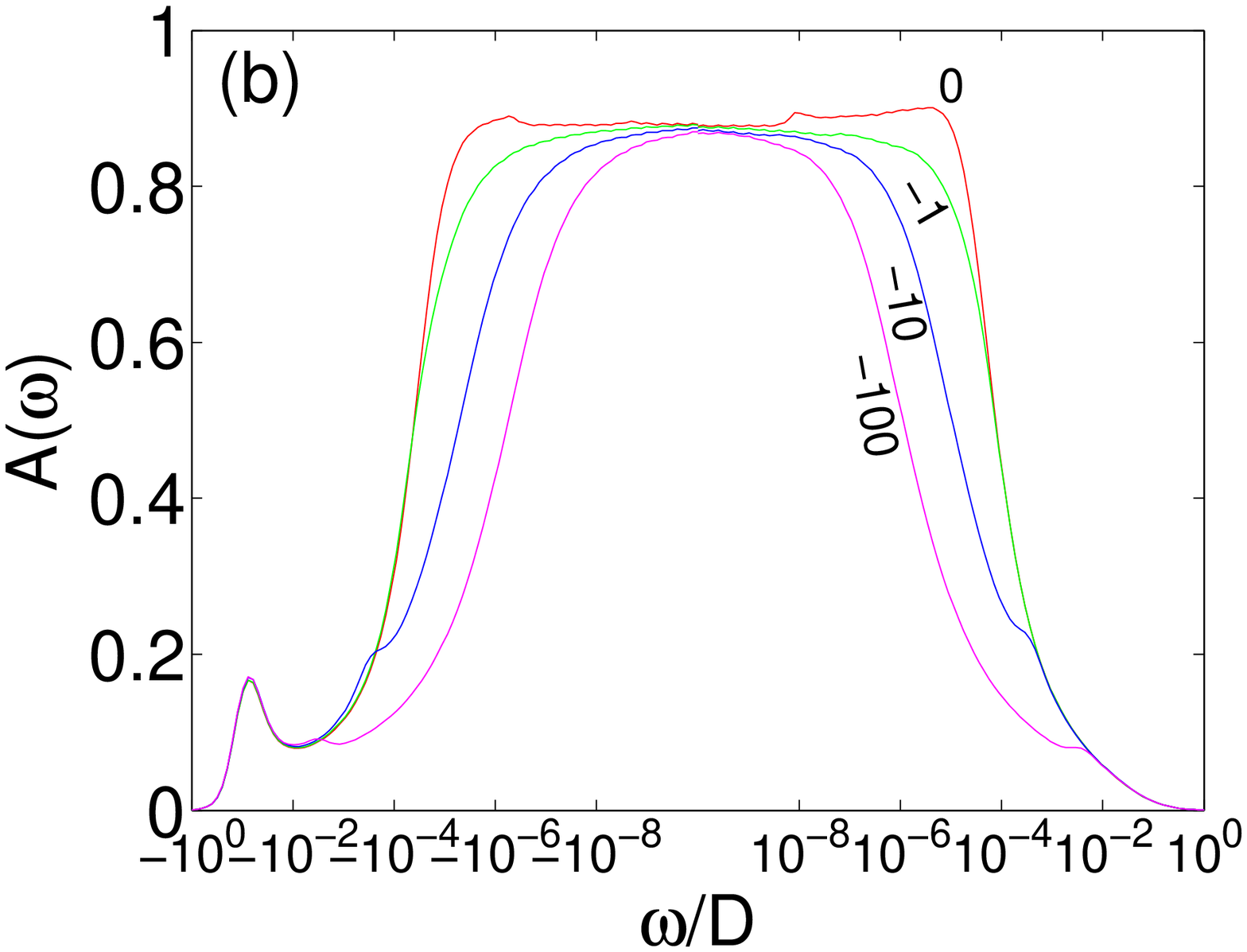}\\
  \includegraphics[width=4.5cm]{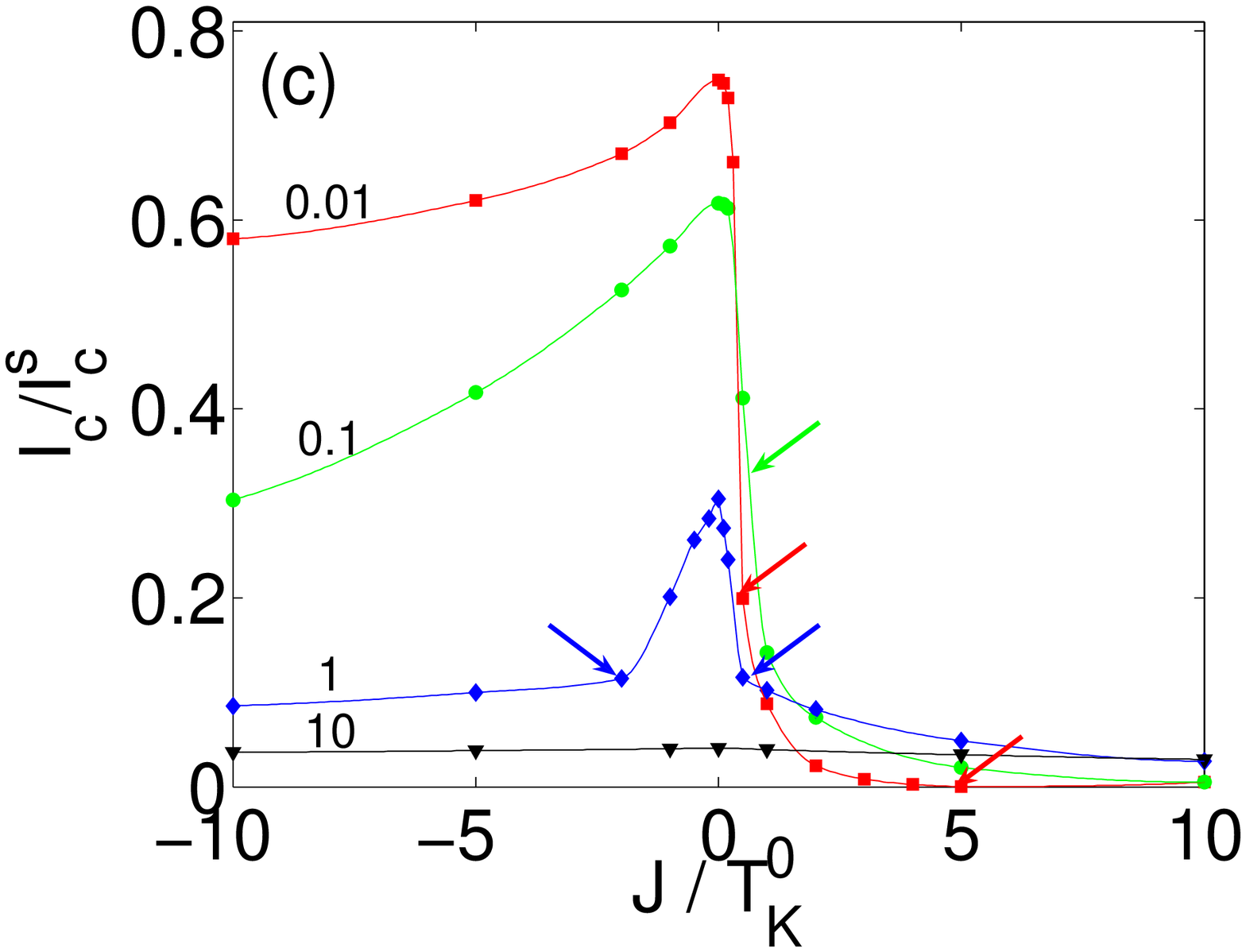}
  \caption{(color online) Spectral weights $\varA(\omega)$ for dot
    electrons coupled to normal leads with antiferromagnetic [(a)] and
    ferromagnetic [(b)] exchange coupling to ion spin for various values of
    $J/T_K^0$ (as annotated). (c) Critical currents as functions of $J/T_K^0$
    for different values of $\Delta/T_K^0$ (see the annotations). The arrows
    locate transition points corresponding to data with the same color. Here we
    have used $S=1/2$.}
  \label{fig:3}
\end{figure}

Finite exchange coupling between electron spins and the ion spin introduces
another electronic correlation and affects Cooper pair
transport. \figref{fig:2} shows typical variations of ALs and supercurrents
with $J$ along the line $aa'$ (see \figref{fig:1}) in the strong coupling limit
($\Delta/T_K^0 = 0.1$). Any finite $J$ clearly induces a splitting in subgap
excitations and consequently causes a crossing between the ground state and the
lowest excitation at $\phi\ne\pi$ (at least for $|J/T_K^0| \lesssim O(1)$); the
level crossing otherwise takes place only at $\phi=\pi$. Across the crossing,
the ground state spin is changed from $S$ to $S{\mp}1/2$ for
$J\gtrless0$. Similarly, the ALs defined as the one-electron/hole subgap
excitations (identified as the poles of the dot Green's functions
\cite{Andreev}) exhibit discontinuities like kinks in the spectra; for
$J\gtrless0$ two outmost ALs with spin $S{\pm}1/2$ with respect to the spin-$S$
ground state cannot remain as (spin-1/2) one-electron excitations with respect
to the ground state with spin $S{\mp}1/2$ at the transition and are replaced by
new ALs with spin $S{\mp}1$. In parallel with an abrupt change in ALs, the
supercurrent-phase relation (SPR) shows a discontinuous sign change (note that
$I \propto - \partial E_A/\partial\phi$, as the continuum-excitation
contribution is negligible \cite{Andreev}), culminating in a transition from 0
to $\pi$ state: two $\pi^{(\prime)}$ states labeled as $\pi_{1,2}^{(\prime)}$
are identified according to the ground-state spin $S\mp1/2$, respectively. The
intermediate states $0'_1$ and $\pi'_{1/2}$ are defined as in Ref.
\cite{Rozhkov}.  The full 0 state exists only at $J=0$ because any small $J$
drives the system to the $\pi$ state at $\phi=\pi$; see \figref{fig:2}.  The
curve of $I(\phi)$ then has three distinct segments \cite{Choi04}. The central
segment resembles that of a short ballistic junction, while the two surrounding
segments are parts of $\pi$-junction curve.  As $J$ grows in magnitude the
central segment shrinks and eventually vanishes. The SPR then becomes
sinusoidal like in a tunnel junction. It should be noted that the 0-$\pi$
transition is asymmetric with respect to the sign of $J$: the transition for
$J>0$ takes place at $\delta E_{\rm S}\sim T_K^0$, where $\delta E_{\rm S} =
\frac{J}{2}(2S{+}1)$ is the exchange-coupling energy gap, while the 0 state
survives much larger ferromagnetic coupling ($J<0$). Once the $\pi$-junction is
fully established, stronger ferromagnetic coupling does not lead to any
qualitative change in the SPR, while a second transition back to 0 state is
observed for large antiferromagnetic coupling $(J\gg\Delta)$. The NRG results
distinguish the second 0 state ($0_2$) from the former one ($0_1$) in three
points: (1) the ground state has spin $S-1/2$ like the $\pi_1$ phase, (2) the
SPR is that of a tunneling junction, and (3) the $\pi_1$-$0_2$ transition has
no intermediate state. \Figref{fig:3}~(c) shows that the critical current has
its maximum at $J=0$ and decreases with increasing $|J|$ rapidly across the
phase boundary for $J>0$ or rather gradually for $J<0$. The critical current
totally vanishes at the $\pi_1$-$0_2$ boundary and increases again slowly with
$J$ in $0_2$ phase (see the curve for $\Delta/T_K^0=0.01$).

The 0-$\pi$ transitions ($0_1$-$\pi_1$ and $0_1$-$\pi_2$) can be attributed to
the competition between superconducting and Kondo correlations as in the
absence of exchange coupling. The relevant parameters are then the Kondo
temperature $T_K$ and the superconducting gap $\Delta$, and the 0-$\pi$ phase
transition occurs when they are comparable to each other: In our choice of
parameters the transition happens at $\Delta_c/T_K \approx 1.84$.  The exchange
coupling manifests itself by renormalizing the Kondo temperature $T_K(J)$. To
see this, we applied the poor man's scaling theory to a corresponding Kondo
Hamiltonian with no superconductivity and $S=1/2$: $\varH_{\rm KM} = \sum_\bfk
\epsilon_\bfk n_\bfk + J \bfS\cdot\bfS_\E + (J_{\rm K} \bfS_\E + J_{\rm M}
\bfS) \cdot\bfS_{\rm L}$, where $\bfS_{\rm L}$ is the spin operator for the
lead electrons at molecule site. The last term $\bfS\cdot\bfS_{\rm L}$
describing direct coupling between spins of the ion and the lead electrons
arises during the scaling process. The renormalization group analysis leads to
the following scaling equations: together with $J \approx J(\Lambda=D)$,
\begin{equation}
    \label{eq:se}
    \ode{J_{\rm K/M}}{\ln\Lambda}
    \approx
    - \rho_0 J_{\rm K/M}^2 + \frac{J}{4D} (2 J_{\rm K} J_{\rm M} - J_{\rm M/K}^2)~.
\end{equation}
As the band width $\Lambda$ is decreased from $D$ to $T_K$, the coefficient
$J_{\rm K}$, responsible for the Kondo correlation, diverges and the scaling
breaks down. In the presence of finite exchange coupling, however, since $J
J_{\rm K} J_{\rm M} > 0$ with $J_{\rm M}(\Lambda=D) = 0$ and $|J_{\rm M}| \ll
J_{\rm K}$, the term proportional to $J$ in \eqnref{eq:se} turns out to slow
down the flow of $J_{\rm K}$ and accordingly lowers the Kondo temperature. This
point is confirmed by NRG calculations applied in the absence of
superconductivity. As can be seen in \figref{fig:3}~(a) and (b), the width of
the spectral density for dot electrons, identified as the Kondo temperature
$T_K(J)$, decreases with increasing $|J|$ (for $J<0$ this decrease, being
marginal, is not clearly shown with the logarithmic scale).  We find out that
for the ferromagnetic case the ratio $T_K(J)/T_K^0$ coincides with
$\Delta_c(J)/\Delta_c(J=0)$. For the antiferromagnetic case, the Kondo
correlation is observed to be suppressed not only by the Kondo peak narrowing
but by lowering the peak height.

Antiferromagnetic exchange coupling can, on the other hand, exert a more
profound effect than simply renormalizing the Kondo temperature: it gives rise
to a reentrant transition to the $\pi$ state at small $\Delta$ and restoration
of the 0 state for large $J$. It is known that small antiferromagnetic exchange
coupling $(J\lesssim T_K^0)$, studied in the context of coupled impurities
\cite{Vojta02} and side-coupled quantum dot systems \cite{SCQD} and observed in
experiments \cite{Roch08}, can produce a two-stage Kondo effect. After the
magnetic moment of the dot is screened by conduction electrons below
$T_K$, at a much lower energy scale (denoted as $T_K^J$) the ion spin is
screened by the local Fermi liquid that is formed on the dot. $T_K^J$ is then
the Kondo temperature of a magnetic moment screened by electrons of a bandwidth
$\sim T_K$ and density of states $\sim 1/(\pi T_K)$ \cite{SCQD}: $T_K^J \sim
T_K \exp\left[-\frac{\pi T_K}{J}\right]$. The second Kondo effect leads to a
Fano resonance and makes a dip in the dot electron density of states as shown
in \figref{fig:3}~(a). The dip becomes widened with $J$ and overrides the Kondo
peak when $T_K^J \approx T_K$ so that the Kondo effect is completely
overridden. As long as $\Delta > T_K^J$, the second Kondo effect does not
appear since the superconducting gap blocks any quasi-particle excitation with
energy less than $\Delta$. For $\Delta\lesssim T_K^J$, however, Cooper pairs
notice the suppression of the Kondo resonance level, and their tunneling is
governed by cotunneling under strong Coulomb interaction, forming a
$\pi$-junction again. Since $T_K^J$ decreases with decreasing $J$, $\Delta_c$
decreases to zero as $J\to0$. Note that the extremely small $T_K^J\ll T_K$
(unless $\delta E_{\rm S}\sim T_K^0$) might make it hard to detect the
reentrance even under rather weak thermal fluctuations with $T_K > T > T_K^J$.

The revival of the 0-state for strong antiferromagnetic coupling can be
explained in the picture of cotunneling of Cooper pairs \cite{Spivak91}. In
weak coupling limit, the fourth-order perturbation theory leads to the
supercurrent:
\begin{align}
  \label{eq:Ip}
  I
  & =
  \frac{4e}{\hbar} \sin\phi
  \sum_{\bfk\bfk'}
  t_{\rm L}^2 t_{\rm R}^2
  \frac{u_\bfk u_{\bfk'} v_\bfk v_{\bfk'}}{\varE_{\bfk} \varE_{\bfk'}}
  \\
  \nonumber
  & \qquad\mbox{}
  \times
  \frac{1}{2S{+}1}
  \left(
    \frac{1}{E_\bfk {+} E_{\bfk'}}
    -
    \frac{2S{+}2}{\delta E_{\rm S} {+} E_\bfk {+} E_{\bfk'}}
  \right),
\end{align}
where $E_\bfk = \sqrt{\Delta^2 + \epsilon_\bfk^2}$, $u_\bfk = \sqrt{(1 +
  \epsilon_\bfk/E_\bfk)/2}$, $v_\bfk = \sqrt{(1 - \epsilon_\bfk/E_\bfk)/2}$,
and $\varE_\bfk = - \epsilon_d - \frac{J}{2}(S+1) - E_\bfk < 0$. For
antiferromagnetic coupling, the ground state for the uncoupled system has spin
$S-1/2$. After one electron tunnels through the molecule the system can be in
spin eigenstate of either $S-1/2$ and $S+1/2$. The latter virtual process,
costing more energy by the gap $\delta E_{\rm S}$, turns out to acquire a $\pi$
phase, contributing to a negative supercurrent. The larger amplitude of this
process by a factor $2S+2$ (degeneracy of the spin state $S+1/2)$ dominates
over spin-preserving process as long as the gap $\delta E_{\rm S}$ is
small. For a large gap $\delta E_{\rm S}$, however, this process becomes
negligible and the sign of the supercurrent is reversed. Note that according to
\eqnref{eq:Ip} the SPR is always sinusoidal and the current should vanish at
the transition, which is also confirmed in our NRG calculations.

\begin{figure}[!t]
  \centering
  \includegraphics[width=7.8cm]{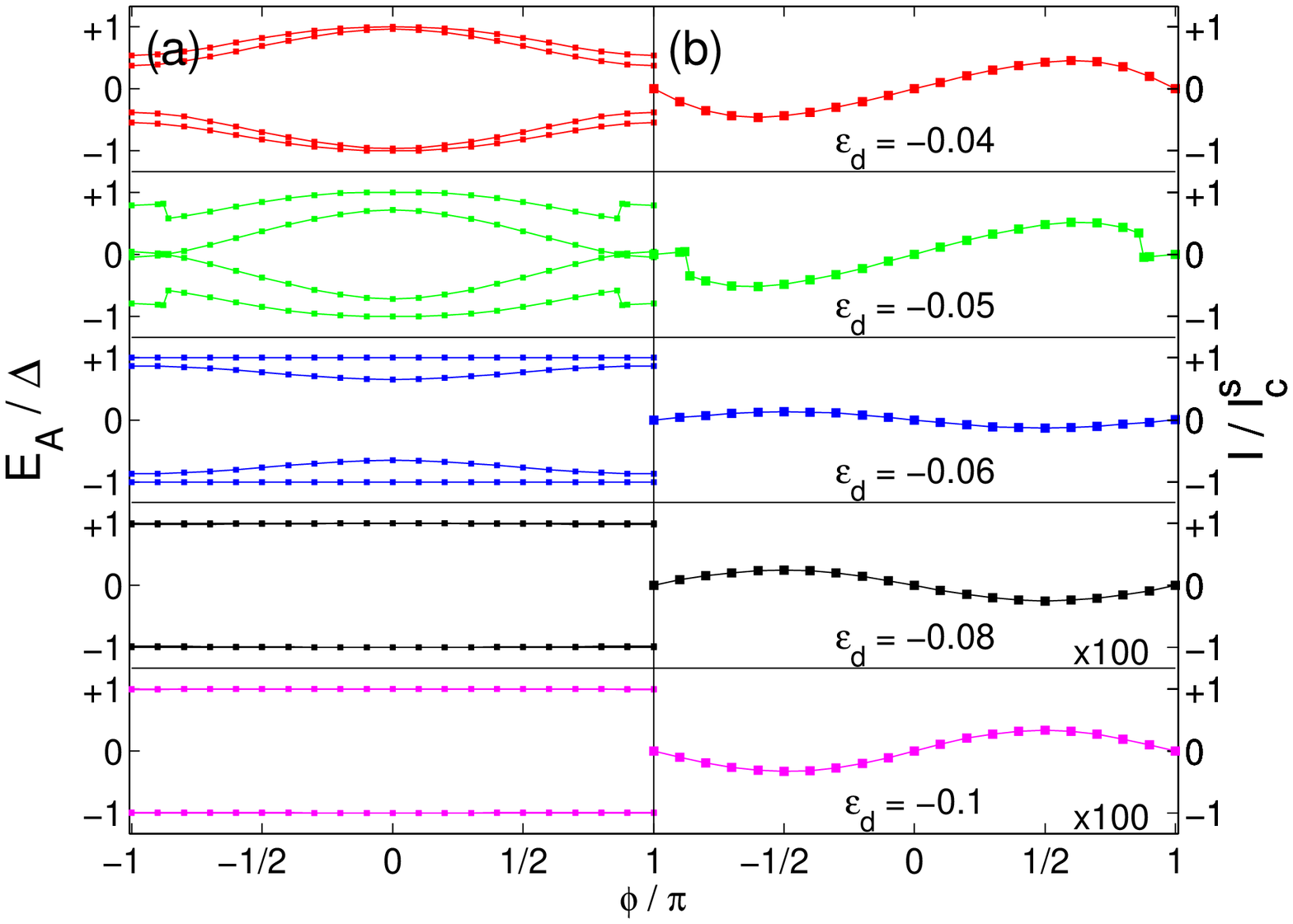}
  \caption{(color online) (a) ALs in units of $\Delta$ and (b) supercurrents
    $I$ in units of $I_c^s$ as functions of $\phi$ with $J/T_K^0 = 10$ and
    $\Delta/T_K^0 = 0.02$ (at $\epsilon_d = -0.1$) while the gate voltage
    $\epsilon_d$ is tuned from $-0.1$ to $-0.04$. See the line $bb'$ in
    \figref{fig:1}.}
  \label{fig:4}
\end{figure}

The physical arguments for the 0-$\pi$ transitions discussed so far are valid
for arbitrary values of the ion spin $S$, while the phase boundaries are
shifted with changing $S$ as shown in \figref{fig:1}. The exchange-coupling
energy gap $\delta E_{\rm S}$ that is supposed to compete with $T_K$ increases
with $S$ so that for larger $S$ the transitions can occur at smaller $J$. On
the other hand, we have observed that the $\pi_1$-$0_2$ transition takes place
at slightly larger $J$ for larger $S$. This is because the increase in the
degeneracy factor $2S+2$ overwhelms the decrease in matrix elements due to a
larger energy cost by $\delta E_{\rm S}$ [see \eqnref{eq:Ip}].

Finally, we present potential experimental manifestations of
exchange-coupling-driven 0-$\pi$ transition. While the direct control of
exchange coupling in molecules is difficult to achieve, the relative strength
$J/T_K^0$ can be controlled by the gate voltage which can tune the Kondo
temperature. \figref{fig:4} proposes a possibility to observe a double
transition (along the line $bb'$ in \figref{fig:1}) as the gate voltage is
swept. Note that the double transition is an evidence of strong exchange
coupling $(J\gg T_K^0\gg\Delta)$: for examples, with $T_K^0\sim3{\rm K}$
measured in a recent C$_{60}$ single-molecular transistor \cite{Roch08}, one
estimates $J\sim30{\rm K}$.  Asymmetry of the phase diagram enables the sign
and possibly the amplitude of $J$ to be determined without ambiguity by
observing the evolution of the SPR or the critical current.

The authors thank W. Wernsdorfer, F. Balestro, and Mahn-Soo Choi for helpful
discussions. This work is supported by ANR-PNANO Contract MolSpintronics
No. ANR-06-NANO-27.


\begin{thebibliography}{99}

\bibitem{Rocha05}
  A. R. Rocha {\it et al.}, 
  Nature Materials \textbf{4}, 335 (2005).

\bibitem{Romeike1}
  C. Romeike {\it et al.}, 
  Phys. Rev. Lett. \textbf{96}, 196601 (2006);
  {\it ibid}. \textbf{97}, 206601 (2006).

\bibitem{Romeike2}
  C. Romeike {\it et al.}, 
  Phys. Rev. Lett. \textbf{96}, 196805 (2006).

\bibitem{Elste}
  F. Elste, and C. Timm, Phys. Rev. B \textbf{71}, 155403 (2005);
  {\it ibid.} \textbf{73}, 235304 (2006);
  {\it ibid.} \textbf{73}, 235305 (2006).

\bibitem{Heersche06}
  H. B. Heersche {\it et al.}, 
  Phys. Rev. Lett. \textbf{96}, 206801 (2006).

\bibitem{Sessoli93}
  R. Sessoli {\it et al.}, 
  Nature (London) \textbf{365}, 141 (1993).

\bibitem{Grose}
  J. E. Grose {\it et al.}, arXiv:0805.2585v1.

\bibitem{Shiba69}
  H. Shiba and T. Soda, Prog. Theor. Phys. \textbf{41}, 25 (1969).

\bibitem{Glazman89}
  L. I. Glazman and K. A. Matveev, Pis'ma Zh. Teor. Fiz. \textbf{49}, 570 (1989) [JETP Lett. \textbf{49}, 659 (1989)].

\bibitem{Spivak91}
  B. I. Spivak and S. A. Kivelson, Phys. Rev. B \textbf{43}, 3740 (1991).

\bibitem{Rozhkov}
  A. V. Rozhkov and Daniel P. Arovas, Phys. Rev. Lett. \textbf{82}, 2788 (1999);
  A. V. Rozhkov and Daniel P. Arovas, Phys. Rev. B \textbf{62}, 6687 (2000);
  A. V. Rozhkov {\it et al.}, Phys. Rev. B \textbf{64}, 233301 (2001).

\bibitem{vanDam06}
  J. A. van Dam {\it et al.}, 
  Nature  (London) \textbf{442}, 667 (2006).

\bibitem{ExpPiJunction}
  V. V. Ryazanov {\it et al.}, 
  Phys. Rev. Lett. \textbf{86}, 2427 (2001);
  T. Kontos {\it et al.}, 
  Phys. Rev. Lett. \textbf{89}, 137007 (2002).

\bibitem{golhaber_gordon}
  D. Goldhaber-Gordon {\it et al.}, 
  Nature (London) \textbf{391}, 156 (1998);
  S. M. Cronenwett {\it et al.}, 
  Science \textbf{281}, 540 (1998).

\bibitem{Choi04}
  M.-S. Choi {\it et al.}, 
  Phys. Rev. B \textbf{70}, 020502 (2004).

\bibitem{Siano04}
  F. Siano and R. Egger, Phys. Rev. Lett. \textbf{93}, 047002 (2004).

\bibitem{Bergeret06}
  F. S. Bergeret {\it et al.}, 
  Phys. Rev. B \textbf{74}, 132505 (2006).

\bibitem{Kasumov05}
  A. Yu. Kasumov {\it et al.}, 
  Phys. Rev. B \textbf{72}, 033414 (2005).

\bibitem{NRG}
  K. G. Wilson, Rev. Mod. Phys. \textbf{47}, 773 (1975);
  H. R. Krishnamurthy {\it et al.}, 
  Phys. Rev. B \textbf{21}, 1003 (1980);
  {\it ibid.}, \textbf{21}, 1044 (1980).

\bibitem{Yoshioka00}
  T. Yoshioka and Y. Ohashi, J. Phys. Soc. Jpn. \textbf{69}, 1812 (2000).

\bibitem{Andreev}
  E. Vecino {\it et al.}, 
  Phys. Rev. B \textbf{68}, 035105 (2003);
  Phys. Rev. Lett. \textbf{91}, 266802 (2003);
  R. L\'opez {\it et al.}, 
  Phys. Rev. B \textbf{75}, 045132 (2007).

\bibitem{Vojta02}
  M. Vojta {\it et al.}, 
  Phys. Rev. B \textbf{65}, 140405(R) (2002).

\bibitem{SCQD}
  P. S. Cornaglia and D. R. Grempel, Phys. Rev. B \textbf{71}, 75305 (2005);
  R. Zitko and J. Bonca, Phys. Rev. B \textbf{73}, 35332 (2006).

\bibitem{Roch08}
  N. Roch {\it et al.}, 
  Nature \textbf{453}, 633 (2008).
\end{thebibliography}
\end{document}